\documentclass[twocolumn,showpacs,preprintnumbers,amsmath,amssymb]{revtex4}
\usepackage[dvips]{graphicx}
\usepackage{dcolumn}
\usepackage{bm}    

\begin{document}
\title{Four-dimensional system with torus attractor birth via saddle-node bifurcation of limit cycles in content of family of blue sky catastrophes}
\author{Alexander~P.~Kuznetsov$^1$, Sergey~P.~Kuznetsov$^1$, \\
and Nataliya~V.~Stankevich$^{2,*}$}
\affiliation{$^{1}$Kotel'nikov's Institute of Radio-Engineering and Electronics of RAS, Saratov Branch,\\
Zelenaya 38, Saratov, 410019, Russian Federation\\
$^{2}$ Yuri Gagarin State Technical University of Saratov, \\
Politechnicheskaya 77, Saratov, 410077, Russian Federation}

\date{\today}
\begin{abstract}
A new four-dimensional model with quasi-periodic dynamics is suggested. The torus attractor originates via the saddle-node bifurcation, which may be regarded as a member of a bifurcation family embracing different types of blue sky catastrophes. Also the torus birth trough the Neimark-Sacker bifurcation occurs in some other region of the parameter space.

$^{*}$Corresponding author. Tel.:+7 8452998818; fax: +7 8452998810

E-mail address: stankevichnv@mail.ru (N.V.Stankevich)
\end{abstract}


\maketitle

\section{Introduction}
Quasi-periodic oscillations represent a wide-spread type of dynamical behavior \cite{p1, p2}. One of the important aspects for understanding these regimes concerns bifurcation scenarios of appearance of the quasiperiodic self-oscillations. Most commonly, one can observe two bifurcation scenarios, which can give rise to a torus attractor, the Neimark-Sacker bifurcation and the saddle-node bifurcation of invariant curves. It seems disappointing that till now a very restricted number of simple model autonomous systems with quasiperiodic dynamics were advanced and studied. In the present work we suggest a four dimensional dynamical system, in which the saddle-node bifurcation of invariant curves producing a two-dimensional torus may be regarded as a member of the bifurcation family of the blue sky catastrophes outlined by Turaev and Shilnikov \cite{p4, pp4}

Originally, the bifurcation of the blue sky catastrophe was described in \cite{p3}. In the simplest case it can be explained as follows. The phase trajectory departs from a vicinity of a semi-stable limit cycle existing at the threshold of the saddle-node bifurcation, goes near a large-size loop along an unstable manifold, and turns back to the limit cycle from the other side. At varying a control parameter in one direction, the semi-stable cycle transforms into a pair of cycles, a stable and an unstable one. If we vary the control parameter in opposite direction, two cycles collide forming the semi-stable cycle, and then disappear, while the large-size limit cycle emerges in the domain containing the helical coils.
According to analysis developed in \cite{p4}, one has to consider, in fact, a family of such bifurcations distinguished by an integer index $m$. Actually, in general, as a phase trajectory of the system departs from the saddle-node cycle attributed with some angular coordinate  $\varphi$, after a travel along the unstable manifold and subsequent return, it will be characterized by the angular coordinate expressed by a relation containing an additive term $m \varphi$. For three-dimensional phase space (minimal dimension where the blue-sky catastrophe may occur) the integer $m$ may be either $0$, or $1$. However, at higher dimensions, any integer can occur. In particular, $m=2$ corresponds to a birth of a hyperbolic strange attractor represented by classic Smale - Williams solenoid.

Conditions and mechanisms of birth of limit cycles through the blue sky catastrophe are described in details in \cite{p4, p5, p6, p7, pp8, pp9}. In \cite{p8} one can find some results demonstrating transition between tonic-spiking and bursting via the blue sky catastrophe in a model of leech neuron. Also, in that paper it was mentioned that this kind of bifurcation can be considered as the main mechanism for the onset of the burst-spike dynamics. In paper \cite{p9} the bifurcation of the blue sky catastrophe has been found in binary mixture contained in a laterally heated cavity at small Prandtl numbers.

In paper \cite{p10} a four-dimensional system was suggested, in which attractor of Smale-Williams type appears as a result of the blue sky catastrophe with the Turaev-Shilnikov index $m=2$. Modifying this system one can easily obtain models with other integer indices $m$ representing various types of the blue-sky catastrophes. In this paper we concentrate on the case $m=1$. It occurs that in this case the bifurcation of collision and disappearance of a pair of small-scale limit cycles leads to appearance of a torus attractor. From a point of view of formal bifurcation theory, this bifurcation does not differ from the saddle-node bifurcation of torus birth observed typically at crossing a border of an Arnold tongue in parameter space of systems manifesting both periodic and quasiperiodic behavior. In our setup, however, this bifurcation appears as a representative of the family embracing the whole assortment of blue-sky catastrophes. It is interesting both for understanding the place of the blue sky catastrophes in the whole picture of bifurcations in dynamical systems and, pragmatically, as an approach to elaboration of concrete realizable systems with definite types of dynamical behavior including quasiperiodic self-oscillations and hyperbolic chaotic attractors.

\section{The basic model and the saddle-node torus bifurcation}

The four-dimensional system with attractor of Smale-Williams type arising via the blue sky catastrophe reads \cite{p10}:
\begin{equation}
\label{eq1}
\begin{array}{l}
  \dot{a_{1}} = -i\omega_{0} a_{1}+(1-|a_{2}|^2+\frac{1}{2} |a_{1}|^2-\frac{1}{50} |a_{1}|^4) a_{1}\\
  \\
  \phantom{\dot{a_{1}} = -i\omega_{0} a_{1}+(1-|a_{2}|^2+\frac{1}{2} |a_{1}|^2)}+\frac{1}{2} \varepsilon \mbox{Im}(a_{2}^2), \\
  \\
  \dot{a_{2}} = -i\omega_{0} a_{2}+(|a_{1}|^2-\mu+\frac{1}{2} |a_{2}|^2-\frac{1}{50} |a_{2}|^4) a_{2}\\
  \\
  \phantom{\dot{a_{1}} = -i\omega_{0} a_{1}+(1-|a_{2}|^2+\frac{1}{2} |a_{1}|^2)}+ \varepsilon \mbox{Re}(a_{1}).
\end{array}
\end{equation}

This system is composed of two coupled self-oscillators with complex amplitudes $a_{1}$ and $a_{2}$; $\omega_{0}$ is the natural frequency of the oscillators, and $\varepsilon$ is the coupling coefficient. If we set the coupling equal to zero ($\varepsilon=0$), and consider the amplitude variables, this system becomes a two-dimensional predator-pray model with some additional parameter $\mu$. This parameter is responsible for bifurcation of equilibrium states that creates conditions for occurrence of the blue sky catastrophe in the full system. The coupling between the subsystems is organized in some special way. In the systems there occurs alternating excitation and dumping of self-oscillatory trains in each oscillator, with transformation of the angular variable representing the phases of the excitation in correspondence with the double-expanding circle map. Due to compression of the phase volume in the state space of the system in all other directions beside the angular variable, attractor of the stroboscopic map is just the Smale-Williams solenoid. In \cite{p10} this mechanism of the system operation is explained in detail.

Let us modify the system (\ref{eq1}) providing linear coupling between the subsystems that makes the equations even simpler:
\begin{equation}
\label{eq2}
\begin{array}{l}
  \dot{a_{1}} = -i\omega_{0} a_{1}+(1-|a_{2}|^2+\frac{1}{2} |a_{1}|^2-\frac{1}{50} |a_{1}|^4) a_{1}\\
  \phantom{\dot{a_{1}} =-i\omega_{0} a_{1}+}+\frac{1}{2} \varepsilon \mbox{Re}(a_{2}), \\
  \dot{a_{2}} = -i\omega_{0} a_{2}+(|a_{1}|^2-\mu+\frac{1}{2} |a_{2}|^2-\frac{1}{50} |a_{2}|^4) a_{2}\\
  \phantom{\dot{a_{2}} = -i\omega_{0} a_{2}+}+ \varepsilon \mbox{Re}(a_{1}).
\end{array}
\end{equation}

Then, the mechanism responsible for formation of the blue sky catastrophe occurs in our system as well; but now it results in appearance of the two-frequency torus attractor instead of the Smale-Williams attractor. The basic two frequencies of the motion on this torus are (i) the natural oscillation frequency $\omega_{0}$, and (ii) that of the long-period oscillatory component, controlled by parameter $\mu$, corresponding to motion around the large-size loop arising in the course of the bifurcation.
\begin{figure}[!ht]
\centerline{
\includegraphics[height=4.5cm, keepaspectratio]{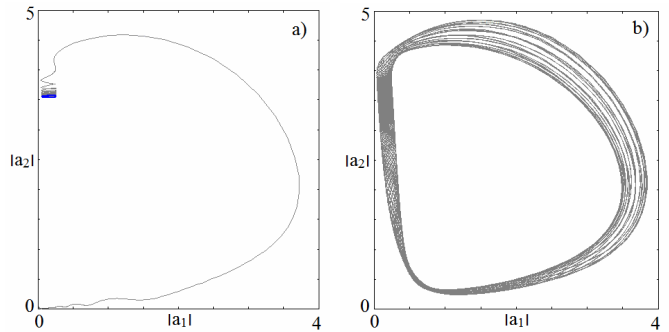}}
\caption{Phase portraits of the system (\ref{eq2}) at $\varepsilon=1$, $\omega_{0}=2\pi$: a) limit cycle and transient process before the torus birth, $\mu=3.19$; b) the torus born as result of the bifurcation similar to the blue sky catastrophe, $\mu=3.2$.}
\label{Fig1}
\end{figure}

Let us turn to the numerical study of the system (\ref{eq2}). In Fig.~\ref{Fig1} the phase portraits are shown in projections on the plane of real amplitudes of coupled oscillators. Panel (a) relates to a subcritical value of $\mu$ just before the bifurcation. The gray thin line corresponds to a transient process. The thick dark line indicates the small-scale stable limit cycle. The phase trajectory arrives there along the unstable manifold of the fixed point at the origin. With increase of the control parameter, the small-scale stable limit cycle meets the small-scale unstable limit cycle (approaching it from below in the used coordinates). After the collision both cycles disappear leaving a channel nearby the domain of their former occurrence, where the orbits pass relatively slowly, returning close to the origin and then back to the input of the channel along the unstable manifold. The result is formation of the large-size attractor represented by a two-frequency torus. Panel (b) shows this torus arisen after the catastrophe due to the collision and disappearance of the stable and unstable small-size limit cycles.

\section{Dynamics of the system in dependence on control parameters}
Let us consider some features of the dynamics of the system in dependence on parameters, and start with the one-parameter analysis.
\begin{figure}[!ht]
\centerline{
\includegraphics[height=7cm, keepaspectratio]{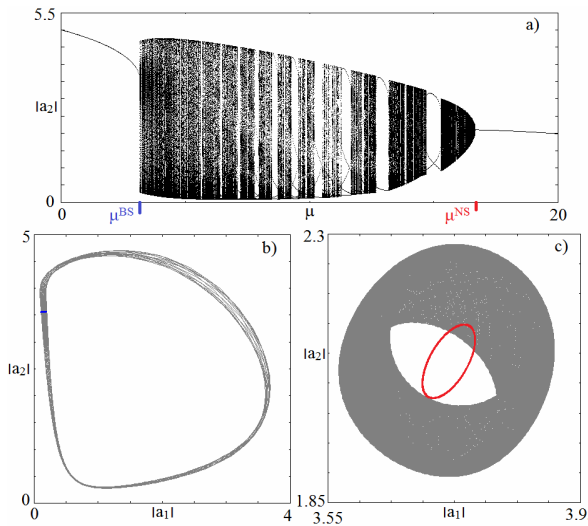}}
\caption{a) Bifurcation diagram of the system (\ref{eq2}) for $\varepsilon=1$, $\omega_{0}=6 \pi$ and phase portraits: b) torus birth via bifurcation similar to the blue sky catastrophe; the limit cycle before bifurcation is marked as bold dark horizontal segment ($\mu=3.17$), and the torus after bifurcation is shown in gray ($\mu=3.18$); c) the torus birth via Neimark-Sacker bifurcation; the bold closed curve is the limit cycle ($\mu=16.7$), and gray color indicates the torus ($\mu=16.65$).}
\label{Fig2}
\end{figure}

In Fig.~\ref{Fig2} panel (a) shows the bifurcation diagram that presents dependence of amplitude of the second subsystem on the control parameter $\mu$. Other parameters are fixed: $\varepsilon=1$, $\omega_{0}=6 \pi$. Two special points are marked in the diagram: $\mu_{c}= \mu^{BS}$ is the critical point of the torus birth as a result of the bifurcation similar to the blue sky catastrophe, and $\mu_{c}=\mu^{NS}$ is the critical point of the torus birth as a result of the Neimark-Sacker bifurcation.
Panels (b) and (c) of Fig.~\ref{Fig2} show phase portraits corresponding to these two critical situations of the torus birth. Bold dark segment in the left diagram indicates the small-size limit cycle just at the bifurcation situation, and gray color designates the new-born torus. Qualitatively, in the diagram one can see clearly that at the blue sky catastrophe, the trajectory which is situated transversally to the small-size limit cycle is stabilized. In panel (c) the limit cycle is shown as bold closed curve, and the torus born via the Neimark-Sacker bifurcation is shown in gray.

As mentioned in \cite{p4, pp4}, in the vicinity of the blue sky catastrophe the characteristic time of returning of the phase trajectory in the Poincar\'{e} section depends on the control parameter as:
\begin{equation}
\label{eq3}
T(\mu)\sim\frac{1}{\sqrt{\mu-\mu_{c}}}.
\end{equation}

\begin{figure}[!ht]
\centerline{
\includegraphics[height=4.5cm, keepaspectratio]{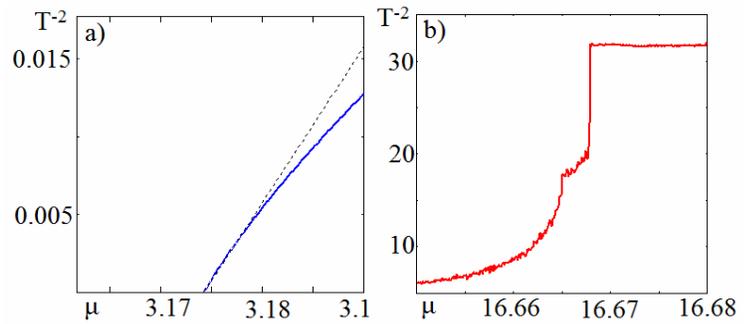}}
\caption{The dependence of squared inverse averaged time of returns at the Poincar\'{e} section on the parameter $\mu$ for $\varepsilon=1$, $\omega_{0}=6\pi$: a) in the vicinity of the bifurcation similar to the blue sky catastrophe, where solid line corresponds to numerical data and dotted line to the approximation (\ref{eq3}); b) in the vicinity of the Neimark-Sacker bifurcation.}
\label{Fig3}
\end{figure}

To check this relation for the system (\ref{eq2}), the returns were calculated nearby the bifurcation point $\mu^{BS}$. The Poincar\'{e} section was produced by a hypersurface defined by the equation $|a_{2}|=c$ with the constant chosen $c=2.1$. In panel (a) of Fig.~\ref{Fig3} we show the plot for the average return times versus $\mu$. As seen, in the vicinity of the critical parameter value $\mu^{BS}$ the dependence for the squared inverse return time is linear. In contrast, nearby the Neimark-Sacker bifurcation $\mu^{NS}$ the dependence looks in absolutely different manner, as seen from the plot on panel (b), and it does not have any pronounced intervals of linearity.

\begin{figure}[!ht]
\centerline{
\includegraphics[height=4.5cm, keepaspectratio]{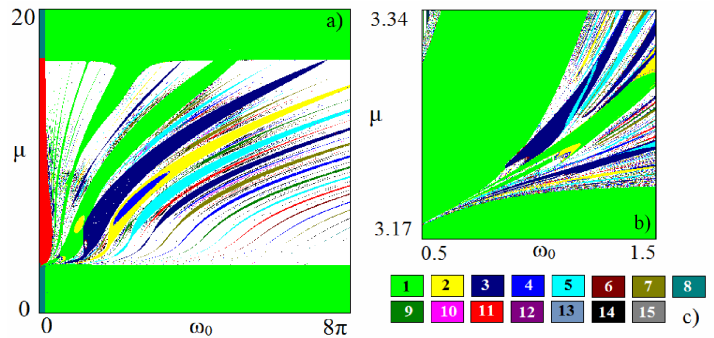}}
\caption{Chart of dynamical regimes a) and its magnified fragment b) for the system (\ref{eq2}) at $\varepsilon=1$; c) palette establishing correspondence of colors and periods of the regimes.}
\label{Fig4}
\end{figure}


Now let us turn to two-parameter analysis of the system (\ref{eq2}). In Fig.~\ref{Fig4} the chart of dynamical regimes is shown obtained numerically for the system (\ref{eq2}) on the parameter plane of the frequency of oscillations $\omega_{0}$ and the coefficient controlling the passage through the blue sky catastrophe $\mu$. This chart was obtained in the following way. We scan the parameter plane with horizontal and vertical steps small enough, and for each point determine in computations a period of the respective sustained regime in the Poincar\'{e} section. (The hypersurface of the Poincar\'{e} section was chosen as $\mbox{Re}(a_{1})=0$.) The largest gray area in panels (a) and (b) correspond to the limit cycle of basic period equal to the period of return to the Poincar\'{e} section. Other periods are multiples of the basic period and are indicated in the parameter plane chart according to the palette shown in panel (c). White areas relate to regimes without recognized finite periods that may represent either quasiperiodic or chaotic oscillations.

In panel (a) one can see two bifurcation lines of disappearance of the limit cycles. As mentioned, the lower line ($\mu\approx3$) corresponds to the bifurcation of torus birth similar to the blue sky catastrophe. The upper line ($\mu\approx16$) corresponds to the Neimark-Sacker bifurcation. One more circumstance confirming the nature of these bifurcations is a characteristic intrinsic structure of the parameter plane in vicinities of the bifurcations.

Along the upper bifurcation line, the set of synchronization tongues is lined up in correspondence to the rational frequency ratios intrinsic to them. Such structure of the parameter plane is typical for situations associated with the Neimark-Sacker bifurcation. In panel (b) a magnified fragment is presented shown a region near the line of the bifurcation similar to the blue sky catastrophe. Observe that synchronization tongues approach the bifurcation line being pulled together in the narrow beam. Analogous parameter plane structures can be observed in different models of neurons, like Hindmarsh-Rose and Sherman models \cite{p11, p12}.

\section{Conclusion}

In the present paper we suggest a new system manifesting both transitions to quasiperiodic self-oscillations via the bifurcation similar to the blue sky catastrophe and via the Neimark-Sacker bifurcation. For this system main characteristics were calculated, like squared inverse averaged time of returns at the Poincar\'{e} section for orbits on the attractor which behaves in essentially different manner for both scenarios of the torus birth. Also, the structure of parameter plane has been revealed for the system.

The proposed system may serve as initial object for construction other models with blue-sky catastrophes generating attractors of various types characterized by different values of the index $m$ in the Turaev - Shilnikov theory \cite{p4, pp4}. This and similar examples may be of interest for design of electronic devices generating quasiperiodic or chaotic signals, as well as for understanding a variety of dynamical behaviors e.g. in neurodynamics.

\bigskip
{\it The research was supported partially by the RFBR grant No 14-02-00085 and by Russian Federation President Program for leading Russian research schools, grant No NSh-1726.2014.2.}

\end{document}